# Exploring quantum non-locality with de Broglie-Bohm trajectories

Ivan P. Christov

Physics Department, Sofia University, 1164 Sofia, Bulgaria

Email: ivan.christov@phys.uni-sofia.bg

**Abstract**

Here in this paper, it is shown how the quantum nonlocality reshapes probability distributions of quantum trajectories in configuration space. By variationally minimizing the ground state energy of helium atom we show that there exists an optimal nonlocal quantum correlation length which also minimizes the mean integrated square error of the smooth trajectory ensemble with respect to the exact many-body wave function. The nonlocal quantum correlation length can be used for studies of both static and driven many-body quantum systems.

PACS: 31.25.-v; 02.70.Ss

## 1. Introduction

In the quantum trajectory based approaches the quantum nonlocality manifests itself as a statistical dependence between the individual trajectories within the Monte Carlo framework. In the Bohmian formulation the origin of the quantum nonlocality is usually identified with the quantum potential which enters a Newtonian-type equation of motion [1], while other approaches use evolving phase space distributions governed by the Liouville equation [2]. Since the many-body wave function $\Psi(\mathbf{R},t)$ resides in configuration space with arguments being the instantaneous coordinates of all electrons $\mathbf{R}=(\mathbf{r}_1,\mathbf{r}_2,...,\mathbf{r}_N)$ the quantum nonlocality can be considered to be an interaction between different trajectories which represent the momentary coordinates of different replicas of the physical particle. This is why the standard quantum Monte Carlo methods use large ensembles of particles (walkers) to calculate the probability distributions of many-body quantum states in configuration space [3]. A new recent time-dependent quantum Monte Carlo (TDQMC) method uses ensembles of both particles and guide waves which evolve concurrently in physical space-time where each particle (walker) with trajectory $\mathbf{r}^k(t)$ samples its own distribution given by the modulus square of the corresponding guide wave $\varphi^k(\mathbf{r},t)$ [4,5]. In this approach each guide wave obeys a separate time-dependent Schrödinger equation (TDSE) in physical space where an effective interaction potential is introduced to account for the local and nonlocal quantum correlations between the particles. From a statistical point of view, such a set of coupled TDSE describes the evolution of a number of probability distributions, in accordance with the standard interpretation of quantum mechanics. The many-body probability distribution given by the trajectory ensemble in configuration space is then considered to be an intersection of these coupled single-body distributions, which translates the idea of quantum nonlocality to the quantum Monte Carlo language [5]. It is important to point out that since in this approach the Monte Carlo walkers are guided by first-order de Broglie-Bohm equations its predictions need not be related to the Bohmian mechanics and its interpretations.

In this paper we calculate of the nonlocal quantum correlation length (NQCL) within the TDQMC framework. As an example system which exhibits strong quantum correlations we consider the ground state and the time evolution of helium atom.

## 2. General theory

In the fixed-nuclei approximation, the system of $N$ electrons is described by the many-body Schrödinger equation:

$$i\hbar\frac{\partial}{\partial t}\Psi(\mathbf{R},t) = -\frac{\hbar^2}{2m}\nabla^2\Psi(\mathbf{R},t) + V(\mathbf{R})\Psi(\mathbf{R},t) \quad , \tag{1}$$

where $\mathbf{R} = (\mathbf{r}_1,...,\mathbf{r}_N)$ is a 3N dimensional vector in configuration space which specifies the coordinates of N electrons, and $\nabla = (\nabla_1,\nabla_2,...,\nabla_N)$. The potential $V(\mathbf{R})$ in Eq. (1) is a sum of electron-nuclear, electron-electron, and external potentials:

$$V(\mathbf{r}_1,...,\mathbf{r}_N) = V_{e-n}(\mathbf{r}_1,...,\mathbf{r}_N) + V_{e-e}(\mathbf{r}_1,...,\mathbf{r}_N) + V_{ext}(\mathbf{r}_1,...,\mathbf{r}_N,t)$$

$$= \sum_{k=1}^{N} V_{e-n}(\mathbf{r}_k) + \sum_{k>l}^{N} V_{e-e}(\mathbf{r}_k - \mathbf{r}_l) + V_{ext}(\mathbf{r}_1,...,\mathbf{r}_N,t) \; . \tag{2}$$

The TDQMC approach assigns classical walkers to each electron degree where the trajectories evolve according to the de Broglie-Bohm guiding equation:

$$\mathbf{v}(\mathbf{r}_i^k) = \frac{\hbar}{m} \mathrm{Im}\left[ \frac{1}{\Psi^k(\mathbf{r}_1,...,\mathbf{r}_N,t)} \nabla_i \Psi^k(\mathbf{r}_1,...,\mathbf{r}_N,t) \right]_{\mathbf{r}_j = \mathbf{r}_j^k(t)} , \tag{3}$$

where i=1,…,N; k=1,…,M denote the electrons and the walkers (replicas) for each electron, respectively. For no spin variables in the Schrödinger equation the many-body wave function can be represented as an anti-symmetrized product (Slater determinant or a sum of Slater determinants) of the individual guide waves:

$$\Psi^k(\mathbf{r}_1,\mathbf{r}_2,...,\mathbf{r}_N,t) = A \prod_{i=1}^{N} \varphi_i^k(\mathbf{r}_i,t) \; . \tag{4}$$

In the TDQMC method many replicas of the trial wave function in Eq. (4) are generated with one walker picked up which belongs to the probability distribution given by each separate guide wave. The set of these walkers represents the probability density of the many-body quantum state in configuration space. The guide waves obey a set of coupled TDSE:

$$i\hbar\frac{\partial}{\partial t}\varphi_i^k(\mathbf{r}_i,t)=\left[-\frac{\hbar^2}{2m}\nabla_i^2+V_{e-n}(\mathbf{r}_i)+\sum_{j\neq i}^{N}V_{e-e}^{eff}[\mathbf{r}_i-\mathbf{r}_j^k(t)]+V_{ext}(\mathbf{r}_i,t)\right]\varphi_i^k(\mathbf{r}_i,t)\,, \tag{5}$$

where the effective electron-electron potential $V_{e-e}^{eff}[\mathbf{r}_i-\mathbf{r}_j^k(t)]$ can be expressed as a Monte Carlo sum over the smoothed walker distribution:

$$V_{e-e}^{eff}[\mathbf{r}_i-\mathbf{r}_j^k(t)]=\frac{1}{Z_j^k}\sum_{l=1}^{M}V_{e-e}[\mathbf{r}_i-\mathbf{r}_j^l(t)]\,K\left(\frac{\left|\mathbf{r}_j^l(t)-\mathbf{r}_j^k(t)\right|}{\sigma_j^k\left(\mathbf{r}_j^k,t\right)}\right), \tag{6}$$

where:

$$Z_j^k=\sum_{l=1}^{M}K\left(\frac{\left|\mathbf{r}_j^l(t)-\mathbf{r}_j^k(t)\right|}{\sigma_j^k\left(\mathbf{r}_j^k,t\right)}\right), \tag{7}$$

is the weighting factor, and $K$ is a smoothing kernel. The main idea behind the nonlocal representation of the effective potential in Eq. (6) is that it entangles the trajectories where the k-th walker from the j-th electron ensemble experiences the Coulomb field of not only the k-th walkers from the ensembles that represent the rest of the electrons, but also due to other walkers from these ensembles which lie within the range of the nonlocal quantum correlation length $\sigma_j^k\left(\mathbf{r}_j^k,t\right)$. Since the Coulomb field in Eq. (6) is smoothed over the contributions of walkers which represent the rest of the electrons, we can relate $\sigma_j^k\left(\mathbf{r}_j^k,t\right)$ to the statistical parameters of the walker ensembles which determine their properties both locally and globally. An especially appropriate smoothing parameter turns to be the kernel density estimation (KDE) bandwidth $\sigma_{j,KDE}^k\left(\mathbf{r}_j^k,t\right)$ which is used to transform discrete to continuous distributions [6]. In general, one can assume that $\sigma_j^k\left(\mathbf{r}_j^k,t\right)$ is a function of $\sigma_{j,KDE}^k\left(\mathbf{r}_j^k,t\right)$:

$$\sigma_j^k\left(\mathbf{r}_j^k,t\right)=F\left[\sigma_{j,KDE}^k\left(\mathbf{r}_j^k,t\right)\right], \tag{8}$$

which, in the most simple linear approximation, becomes:

$$\sigma_j^k\left(\mathbf{r}_j^k,t,\alpha_j\right)=\alpha_j.\sigma_{j,KDE}^k\left(\mathbf{r}_j^k,t\right), \tag{9}$$

where the parameters $\alpha_j$ can be determined by variationally minimizing the ground state energy of the quantum system [5]. Here, we assume that $\alpha_j$ is the same for all electrons, and denote it by $\alpha$. It is important to point out that for $\alpha\to 0$ the effective potential in Eq. (6) tends to the pairwise *e*-*e* potential, while for $\alpha\to\infty$ the effective potential reduces to the mean-field (Hartree-Fock) potential. In the intermediate case the *k*-th walker from the *i*-th electron ensemble would experience the full Coulomb potential due to the *k*-th walker form the *j*-th electron ensemble and the weighted Coulomb potentials due to the rest of the walkers which represent the *j*-th electron, which is a direct manifestation of the quantum nonlocality.

## 3. Calculation of the nonlocal quantum correlation length

In order to explore the linear approximation to the nonlocal quantum correlation length given by Eq. (9) we fist calculate the ground state of a strongly correlated model system (one-dimensional helium atom). This model atom has proven to be very useful in modeling the interaction of atomic systems with intense ultrashort laser pulses (e.g. in [7]) where modified Coulomb potentials have been employed to avoid numerical complications from the singularity at the origin. Additional advantage of this model is that the corresponding two-body time-dependent Schrödinger equation can be solved numerically very accurately. Here we assume that the electron-nuclear and the electron-electron interactions are approximated by the following potentials:

$$V_{e-n}(x_i) = -\frac{2e^2}{\sqrt{a + x_i^2}}; \qquad (10)$$

$$V_{e-e}[x_i - x_j^k(t)] = \frac{e^2}{b + \left| x_i - x_j^k(t) \right|}, \qquad (11)$$

where $i$=1,2; $k$=1,…,M, and we have chosen $a$=$b$=1 a.u. (atomic units) in Eqs. (10), (11). The ground state of the atom is prepared as described previously [4,5]. First a separate guiding wave $\varphi_i^k(x_i, t=0) = \exp\left(-x_i^2\right)$ is assigned to each walker $x_i^k(t=0)$ from initial Monte Carlo ensembles of M=25 000 particles and guide waves for the two electrons. Next, both waves and walkers are propagated over 400 complex time steps in the presence of a random component that thermalizes the ensemble to avoid possible bias in the walker distribution that may arise due to the quantum drift alone. Each separate walker samples its guiding wave's distribution using Metropolis algorithm. The complex time with equal real and imaginary parts ensures a balanced nonzero velocity of the walkers for evolution towards stationary state.

In general, the NQCL which determines the characteristic dimensions of coupling between the walkers from ensembles that belong to different electrons is a position-dependent quantity, as is the KDE bandwidth (see Eq. (9)). However, for clarity here we consider the constant bandwidth approximation which implies also constant correlation length over the whole Monte Carlo ensemble which represents a given electron degree. In this case, Eq. (9) reduces to:

$$\sigma_j^k\left(t,\alpha\right)=\alpha.\sigma_{j,KDE}^k\left(t\right) \tag{12}$$

It is clear that even for a constant (in space) correlation length significant reshaping of the Monte Carlo ensembles may occur because the correlations modify the Coulomb repulsion experience by each walker locally. In order to visualize the effect of reshaping we plot with blue contour lines in Fig. (1) the smoothed Monte Carlo distributions as compared to the density obtained from the numerical solution of the two-dimensional Schrödinger equation for the ground state (red contours), which we will call henceforth "exact" density. The smoothing is performed using Gaussian kernel function $K\left(x\right)=\exp\left[-x^2/\left(2\sigma_{KDE}^2\right)\right]$ in Eqs. (6), (7). Figures 1 (a) and 1 (c) depict the limiting cases where $\alpha\to 0$ and $\alpha\to\infty$, respectively. The characteristic butterfly shape of the blue contours in Fig. 1 (a) with dents along the diagonal evidences the effect of the electron repulsion for the ultra-correlated (zero correlation length) case while the mean field (Hartree-Fock) distribution in Fig. 1 (c) is closer to square-shaped. Therefore we can interpret the trajectories for $\alpha \to 0$ as ultra-correlated but not entangled, while $\alpha \to \infty$ corresponds to uncorrelated but ultra-entangled trajectories. Figure 1 (b) shows the case where the Monte Carlo distribution is most close to the exact probability density $P\left(\mathbf{R},t\right)$.

In order to find the value for the parameter $\alpha$ which corresponds to the optimal distribution in Fig. 1 (b) we used a variational approach where $\alpha$ is changed while monitoring the ground state energy of the atom given by:

$$E\left(t,\alpha\right)=\frac{1}{M}\sum_{k=1}^{M}\left[\frac{1}{8}\sum_{i=1}^{N}\left.\frac{\left[\nabla_{\mathbf{r}_i}\hat{P}\left(\mathbf{R},t,\alpha\right)\right]^2}{\hat{P}\left(\mathbf{R},t,\alpha\right)^2}\right|_{\mathbf{r}_i=\mathbf{r}_i^k(t)} + \sum_{i=1}^{N}V_{e-n}(\mathbf{r}_i^k)\Big|_{\mathbf{r}_i^k=\mathbf{r}_i^k(t)} + \sum_{i>j}^{N}V_{e-e}(\mathbf{r}_i^k-\mathbf{r}_j^k)\Big|_{\mathbf{r}_{i,j}^k=\mathbf{r}_{i,j}^k(t)}\right], \tag{13}$$

and the mean integrated squared error (MISE) defined as an expectation with respect to several data samples [8]:

$$MISE(t,\alpha)=E\int\left[\hat{P}\left(\mathbf{R},t,\alpha\right)-P\left(\mathbf{R},t\right)\right]^2 d\mathbf{R}, \tag{14}$$

where $\hat{P}\left(\mathbf{R},t,\alpha\right)$ is the kernel density estimate which in our case is expressed in terms of a product kernel:

$$\hat{P}(\mathbf{R},t,\alpha)=\frac{1}{M}\sum_{k=1}^{M}\left\{\prod_{i=1}^{N}\prod_{d=1}^{D}\frac{1}{\sqrt{2\pi}\sigma_{i_d,KDE}^{k}(t,\alpha)}\exp\left[-\frac{\left|\mathbf{r}_{i_d}(t)-\mathbf{r}_{i_d}^{k}(t)\right|^2}{2\sigma_{i_d,KDE}^{k}(t,\alpha)^2}\right]\right\}, \qquad (15)$$

where the index d=1,2…D denotes the axes in physical space of dimension D. The result from the variation of $\alpha$ is plotted in Fig. 2 as a dependence of the ground state energy and of the MISE on $\sigma_j^k(\tau,\alpha)$, where $t=\tau$ is the moment where steady state is established. It is seen that both the ground state energy and the MISE experience minima for the same value of $\sigma_j^k(\tau,\alpha)\approx 1.35$ a.u. The minimum ground state energy in Fig. 2 (a) is -2.4 a.u. in close correspondence with the exact value of -2.399 a.u., while the minimal MISE in Fig. 2 (b) is ~ 4 $10^{-4}$. For this value of $\sigma_j^k(\tau,\alpha)$ there is an almost perfect match between the smooth Monte Carlo data and the exact probability density, as it is seen from Fig. 1 (b). For larger values of $\sigma_j^k(\tau,\alpha)$ the ground state energy approaches the Hartree-Fock value of -2.389 a.u.

Once the optimum $\sigma_j^k(\tau,\alpha)$ for the ground state is found it is interesting to observe the deviations of both NQCL and MISE for real-time propagation of the Monte Carlo ensembles for an atom exposed to a strong external field which can cause significant deformations of the electron cloud. Figure 3 (a) shows the time profile of a few period linearly polarized electromagnetic field $E(t)=E_0(t)\cos(\omega t)$ with peak amplitude 0.15 a.u. and carrier frequency 0.153 a.u. used in the calculation. Figure 3 (b) and Fig. 3 (c) depict the time dependent NQCL and MISE, respectively, for the same value of the parameter $\alpha$ which minimizes the ground state energy. It is seen from Fig. 3

(b) that the NQCL increases by a factor of three for a few periods close to the peak of the pulse where the external field causes significant portions of the electron cloud to leave the atom due to tunneling ionization. At the same time Figure 3 (c) shows that the MISE (which is calculated with respect to the evolving exact solution) remains of the order of $10^{-3}$ , which indicates that the value of the coefficient $\alpha$ we found for the ground state provides a good approximation also for the strong ionization regime. This is further confirmed by Fig. 4 which shows the time dependence of the survival probability for the helium ground state calculated as the portion of the MC walkers for the two electrons which remains within 10 a.u. from the core. It is seen that the results obtained from the time-dependent mean field approximation ($\alpha \to \infty$), and those from the correlated TDQMC calculations, are very close to the corresponding exact curves. As expected, the survival probability for the correlated case in Fig. 4 is lower due to the electron-electron repulsion. We have also verified that the variational approach used here can be applied to find the ground state energy as a function of the nonlocal correlation length in three spatial dimensions [9]. The result for the ground state energy of a 3D helium atom as a function of the NQCL is shown in Fig. 5. It is seen that for $\sigma_j^k(\tau,\alpha) \approx 1.5$ a.u. the energy of the ground state approaches -2.9 a.u., in close correspondence with the exact result.

## 4. Conclusions

In this paper, we have explored the role of the quantum nonlocality for reshaping the many-body quantum distributions in configuration space within the frames of time-dependent quantum Monte Carlo approach. By solving large sets of coupled time dependent Schrödinger equations together with first order guiding equations for the Monte Carlo walkers we variationally optimize the ground state energy and the mean integrated squared error with respect to the nonlocal quantum correlation length. Our calculations reveal that the nonlocal quantum correlation length can be approximated with good accuracy as linearly proportional to the kernel density estimation bandwidth for the Monte Carlo data. For the optimal value of the NQCL, the MISE exhibits a minimum which is found to be quite small ($\sim 10^{-4}$), which indicates a very good fit between the Monte Carlo probability distribution and the exact one. For the minimum value of the MISE the ground state energy calculated from the Monte Carlo data also reaches minimum, in close correspondence with the exact value. This can be used to determine the NQCL for practical cases where the exact ground state is unknown. The accuracy can further be improved by assuming local dependence of the NQCL, e.g. by using adaptive techniques for kernel density estimation (that may include full covariance matrices). In a typical TDQMC calculation the variational optimization of the ground state with respect to the nonlocal correlation length should precede the real time propagation, which can be applied also to constituents which include classical degrees, e.g. molecules, clusters, etc. On the other hand, for many-electron systems (electron gas, metals, etc.) the combination of Eq. (3) and Eq. (4) may become somewhat impractical because of the huge Slater determinants involved. In such cases the well known Slater or Dirac local exchange potentials can be added in Eq.(5), while each walker is being guided by its own guide wave.

The TDQMC method presented here is ideally suited for parallel implementation where communication between the different processes is needed only for the calculation of the nonlocal quantum correlation effects. Our calculations reveal that the TDQMC method offers a controlled-accuracy solution of the quantum many-body problem which scales in time as a low-order polynomial with the number of electrons involved, for up to three spatial dimensions, in contrast to the exact solution which scales exponentially with the system dimensionality. Also, it was found that at least 10 000 Monte Carlo walkers have to be involved in the calculation for accuracy of about 1% from the ground state energy. The calculations presented here were performed using a massively parallel Blue Gene /P supercomputer and SGI Octane III personal supercomputer, with almost the same scaling observed.

## Acknowledgments

The author gratefully acknowledges support from the National Science Fund of Bulgaria under Grant DCVP 02/1 (SuperCA++) and Grant DO-02-114-2008. Computer resources from the National Supercomputer Center (Sofia) are gratefully appreciated.

**Figure captions:**

**Figure 1**. Probability density distributions in configuration space for 1D helium atom: (a)- ultra-correlated case, $\alpha \to 0$; (b)- TDQMC result, $\alpha$ is data determined; (c)- mean field (Hartree-Fock) result, $\alpha \to \infty$. Blue contours – smoothed Monte Carlo data, red contours – exact result.

**Figure 2.** Ground state energy (a) and mean integrated squared error (b) as function of the nonlocal quantum correlation length for 1D helium atom.

**Figure 3**. Time dependence of the electric field (a), the nonlocal quantum correlation length (b), and the mean integrated squared error (c) for 1D helium exposed to a strong laser pulse.

**Figure 4.** Time-dependent survival probability for 1D helium atom: blue lines – TDQMC calculation; red lines – exact result.

**Figure 5.** Ground state energy as a function of the nonlocal quantum correlation length for 3D helium atom.

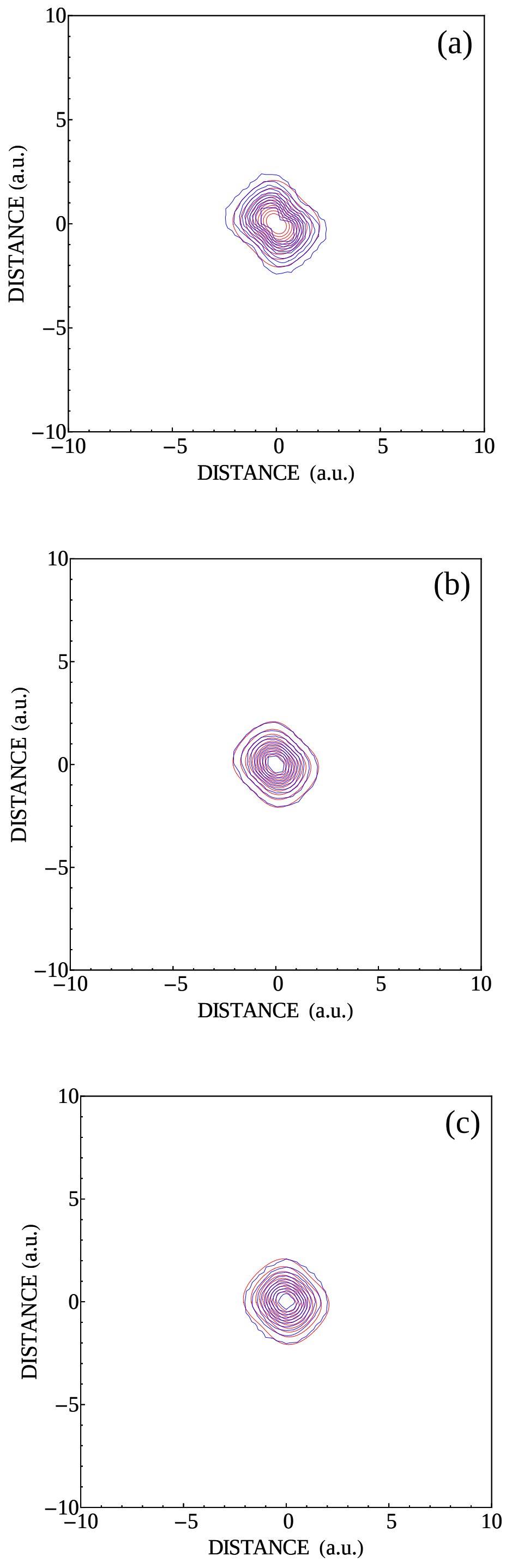

Christov, Figure 1

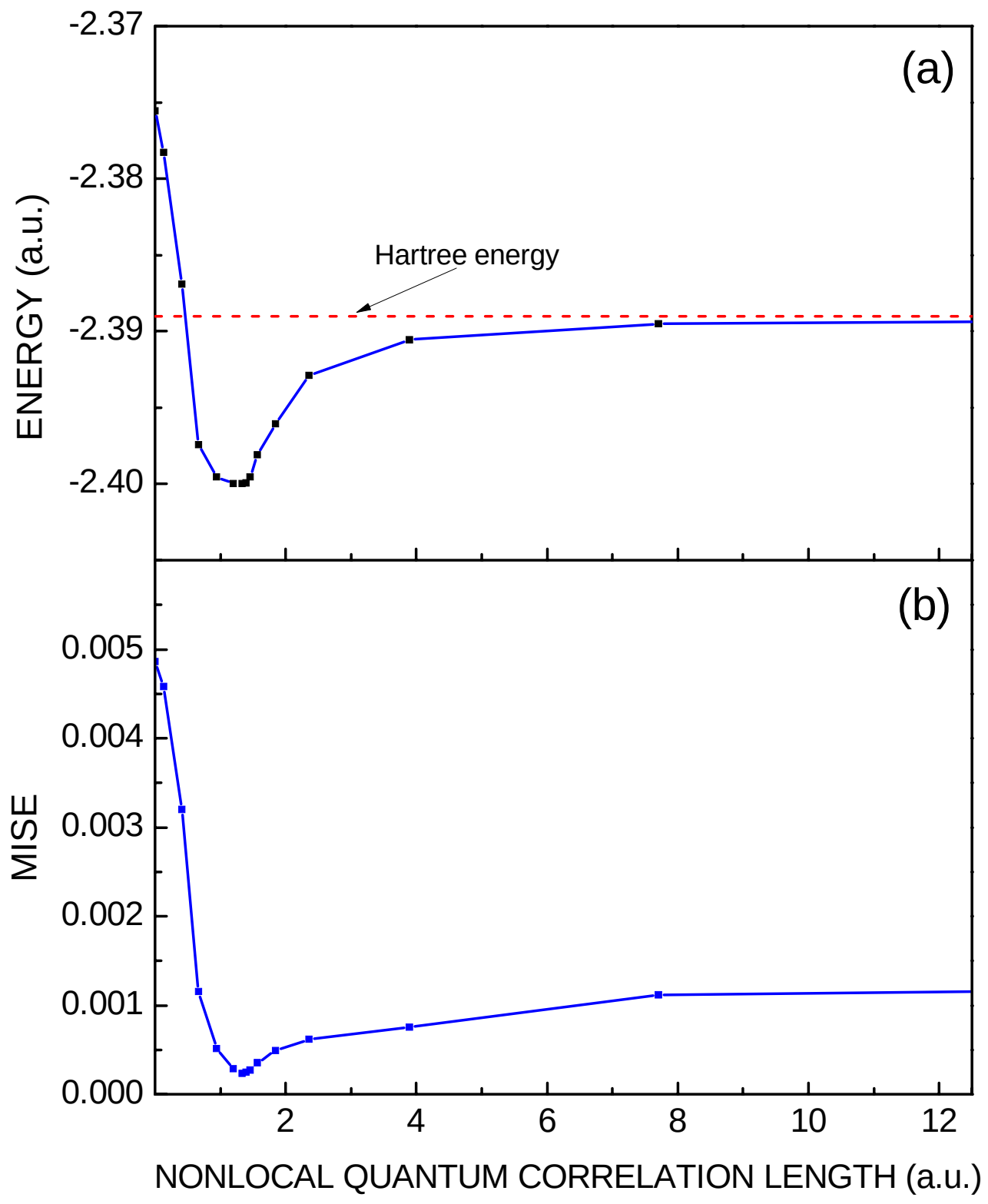

Christov, Figure 2

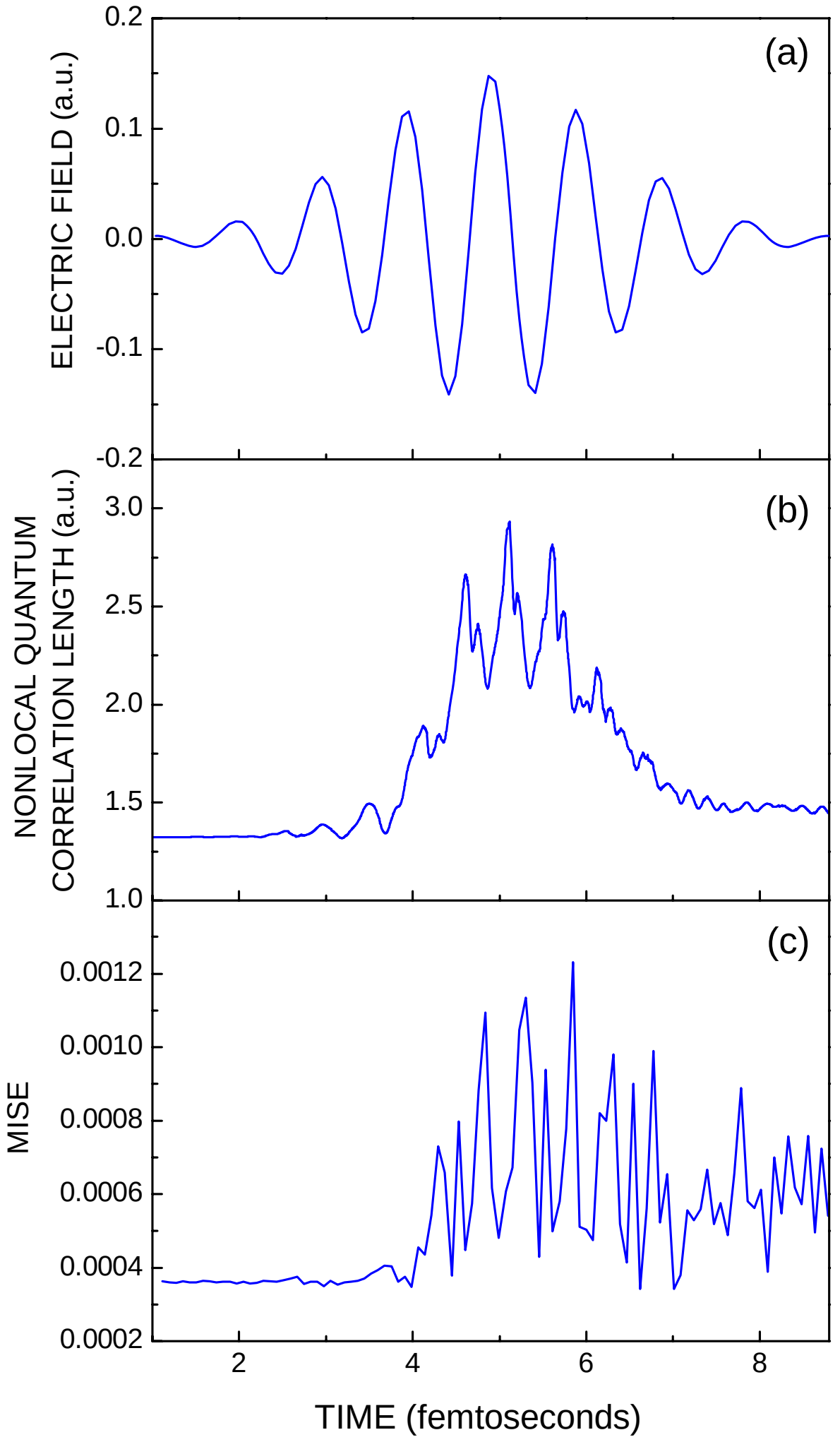

Christov, Figure 3

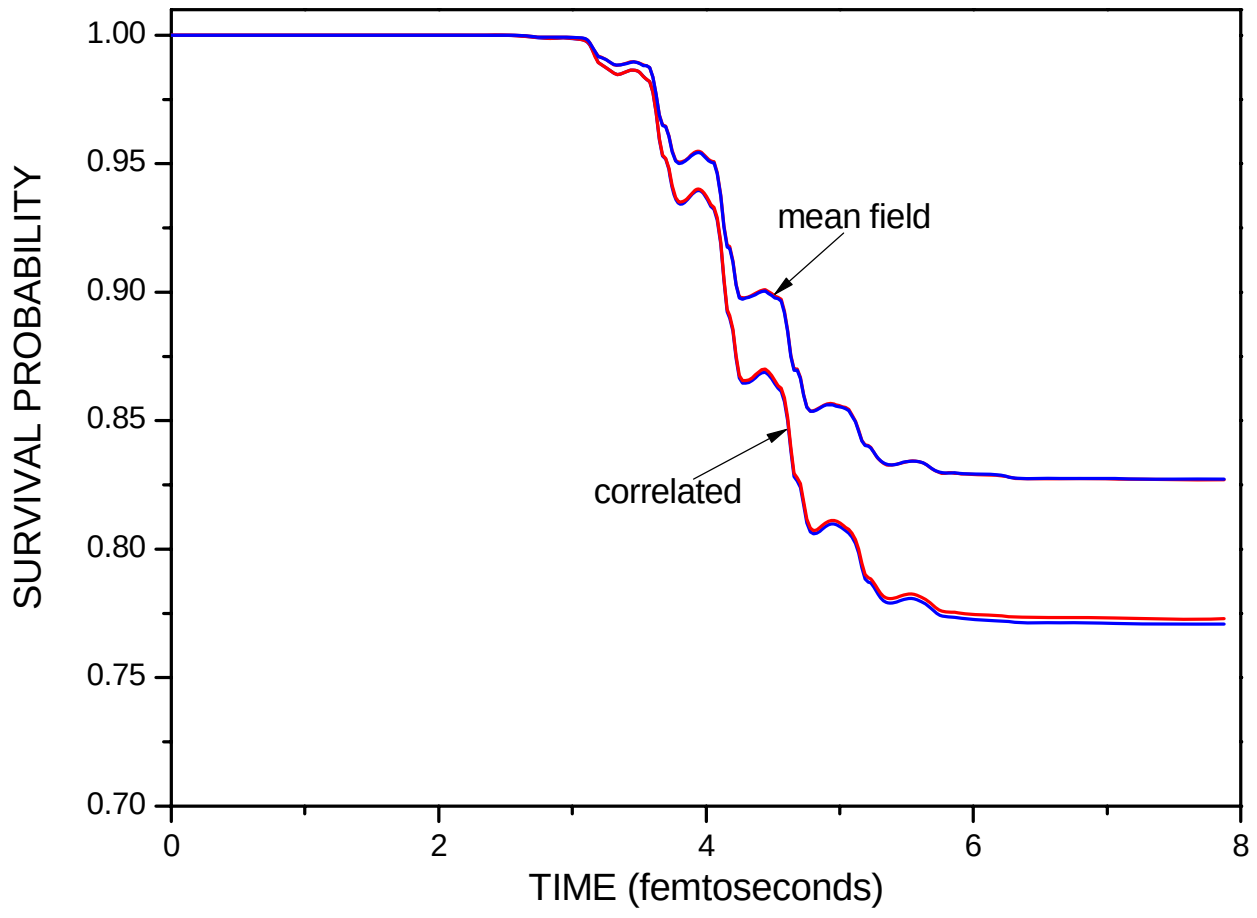

Christov, Figure 4

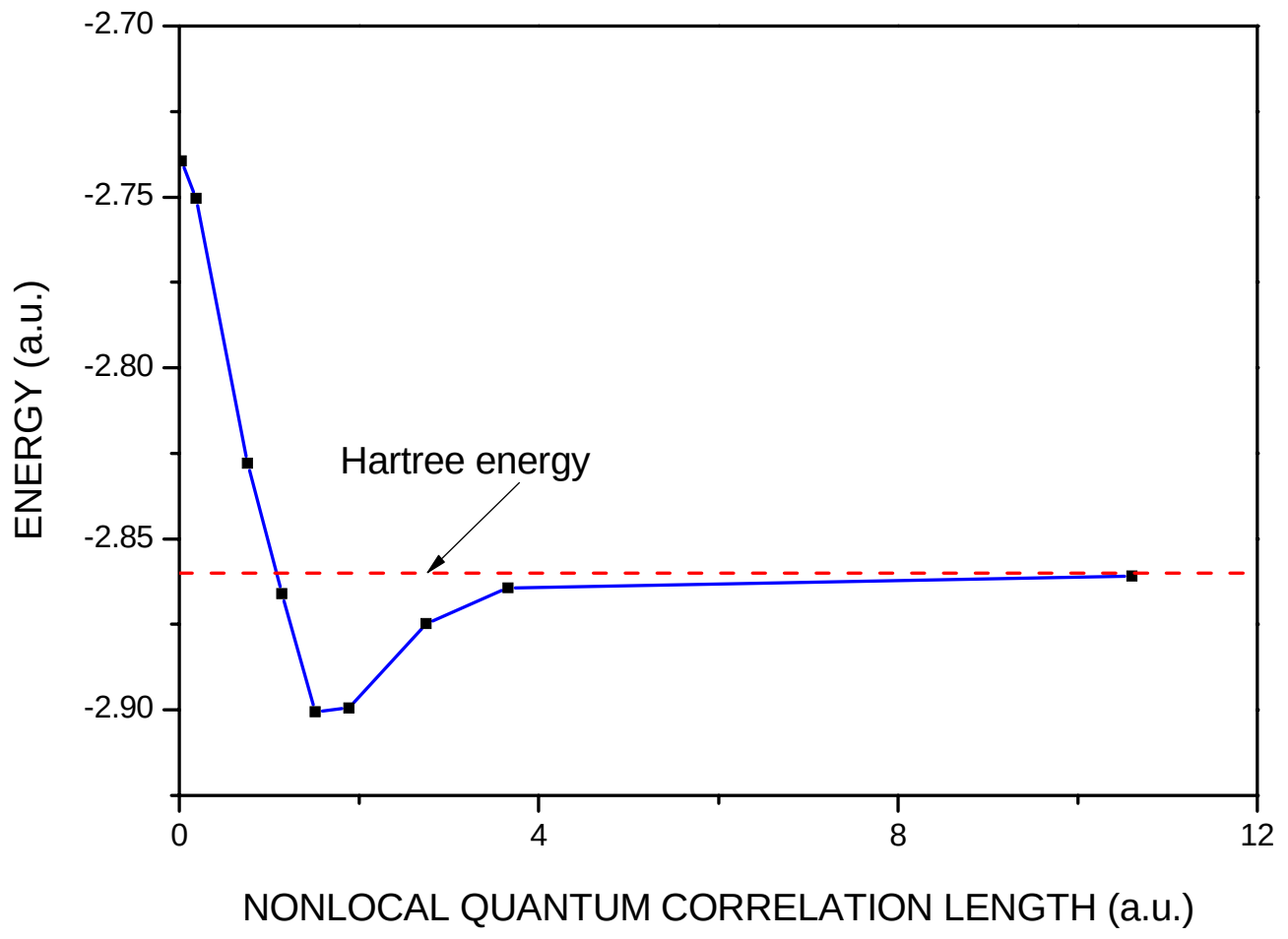

Christov, Figure 5